\documentclass{emulateapj}

\usepackage{amsmath}
\usepackage{cases}
\usepackage{txfonts}
\usepackage{tipa}
\usepackage{amsfonts}
\usepackage{amssymb}
\usepackage{graphicx}
\usepackage{multirow}

\shorttitle{Monte-Carlo Method}
\shortauthors{Gu et al.}

\begin{document}

\title{A Monte-Carlo Method for Estimating Stellar Photometric Metallicity Distributions}

\author{Jiayin Gu\altaffilmark{1}, Cuihua Du\altaffilmark{2}, Yingjie Jing\altaffilmark{2} and Wenbo Zuo\altaffilmark{2}}
\affil{$^{1}$Department of Physics, Wuhan University of Technology, Wuhan 430000, P. R. China; gujiayin12@mails.ucas.ac.cn\\
$^{2}$School of Physical Sciences, University of Chinese Academy of
Sciences, Beijing 100049, P. R. China; ducuihua@ucas.ac.cn}


\begin{abstract}
Based on the Sloan Digital Sky Survey (SDSS), we develop a new monte-carlo based
method to estimate the photometric metallicity distribution function (MDF) for stars in the Milky Way. Compared with other photometric calibration
methods, this method enables a more reliable determination of the MDF, in particular at the metal-poor and metal-rich ends.
We present a comparison of our new method with a previous polynomial-based approach, and demonstrate its superiority. As
an example, we apply this method to main-sequence stars with $0.2<g-r<0.6$, $6$ kpc$<R<9$ kpc, and in different
intervals in height above the plane, $|Z|$. The MDFs for the selected stars within
two relatively local intervals ($0.8$ kpc$<|Z|<1.2$ kpc, $1.5$ kpc$<|Z|<2.5$ kpc) can be well-fit by two Gaussians,
with peaks at [Fe/H] $\approx-0.6$ and $-1.2$ respectively, one associated with the disk system, the other with the halo.
The MDFs for the selected stars within two more distant intervals ($3$ kpc$<|Z|<5$ kpc, $6$ kpc$<|Z|<9$ kpc) can be decomposed into three Gaussians,
with peaks at [Fe/H] $\approx-0.6$, $-1.4$ and $-1.9$ respectively, where the two lower peaks may provide evidence for a two-component model of the halo: the inner halo and the outer halo.
The number ratio between the disk component and halo component(s) decreases with vertical distance from the Galactic plane,
consistent with the previous literature.

\end{abstract}


\section{Introduction}

\par The metallicity distribution function (MDF) for stars in the Milky Way is of great importance
to reveal the chemical structure of Galactic halo and disk systems, and
provides essential clues to the assembly and enrichment history of
the Galaxy \citep[][]{Carollo07, Carollo10, Peng12, Peng13, An13, An15}. Many multi-fiber
spectroscopic surveys, including the Sloan Digital Sky Survey
\citep[SDSS;][]{York00}, the Radial Velocity Experiment
\citep[RAVE;][]{Steinmetz06}, and the Large Sky Area Multi-Object Fiber
Spectroscopic Telescope \citep[LAMOST;][]{Deng12, Liu14, Zhao12}, have
obtained metallicity estimates and other stellar parameters for millions of
stars from low- and medium-resolution spectra.  However, compared
with photometric data, the number of spectra within the limiting
magnitudes of the surveys is still too small to provide a detailed chemical map of the Galaxy,
even in the relatively local region. The use of photometric data,
which is available for far more stars than the spectroscopic data, breaks through this limitation.

\par Considering that
the exhaustion of metals in a stellar atmosphere has a detectable
effect on the emergent flux \citep[][]{Schwarzschild55}, in
particular in the blue region where the density of metal absorption
is highest, the combination of spectroscopic data and photometric
data can be used to derive estimates of [Fe/H] \citep[][]{Allende06,
Allende08, Lee08a, Lee08b}. For example, \cite{Siegel09} derived
approximate stellar metallicities through measurement of the
ultraviolet excess, based on $UBV$ data in SA 141. \cite{Ivezic08}
employed SDSS data to derive a metallicity estimator from $u-g$ and
$g-r$ colors, and successfully mapped the metallicity distribution
of millions of F/G main-sequence stars within a distance of $\sim 8$
kpc from the Sun. \cite{Peng12} also used BATC survey data to estimate the
stellar photometric metallicity distribution. \cite{Gu15}
obtained a metallicity estimator using SCUSS \citep[][]{Zou15,
Zou16} data, which can be applied up to fainter magnitudes due to the
use of more accurate SCUSS $u$-band measurements, and used it to
explore the metallicity of the Sagittarius stream in the South
Galactic cap. Using a minimum $\chi^{2}$ technique, \cite{Yuan15}
estimated photometric metallicities simultaneously using the
dereddened colors $u-g$, $g-r$, $r-i$, and $i-z$ from the SDSS and
metallicity-dependent stellar loci. \cite{An13} calibrated stellar
isochrones to derive metallicities of individual stars with SDSS
$ugriz$ photometry. An et al. (2015) applied this method to
recently improved co-adds of $ugriz$ photometry for Stripe 82 from
SDSS, including a factor of two more stars than their previous
effort. The new analysis revealed a MDF
for halo stars between 5 and 10 kpc from the Sun with peak
metallicities at [Fe/H] $\sim -1.4$ and [Fe/H] $\sim -1.9$, which
the authors associated with the inner-halo and outer-halo
populations of the Milky Way, respectively.

\par The photometric metallicity calibrations developed by previous works are
characterized by their assignment of a star-by-star metallicity estimate based on its color indexes. This
inevitably introduces error, because a single star's metallicity is actually uncertain even when its color indexes are fixed,
varying around the metallicity estimate to form a distribution. In addition, the calibration methods
used by \cite{Ivezic08} and \cite{Gu15} yield poor results for
very metal-rich or very metal-poor stars. However, in order to
investigate the chemical structure of the Galactic stellar populations, we
only require knowledge of the MDF for a
large statistical sample of stars. Here we develop a monte-carlo method to estimate the photometric metallicity
distribution of large number of stars with available SDSS photometry.

\par This paper is organized as follows. In Section 2,
we provide a brief overview of the SDSS and its photometric data. Details of our monte-carlo based photometric metallicity calibration
are presented in Section 3. Section 4 presents a comparison
between this method and the more traditional polynomial-fitting method. As
an example, in Section 5 we apply this method to derive the MDF for Galactic stars and consider its variation with height above the plane. A brief summary
is given in Section 6.

\section{SDSS photometric data}

\par
\begin{figure}
\includegraphics[width=1.0\hsize]{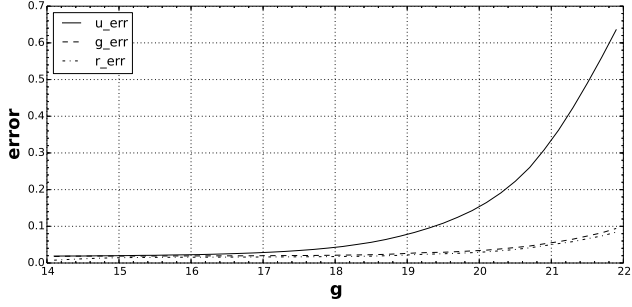}
\caption{The average magnitude error of numerous stars as a function
of $g$ magnitude. Main-sequence stars with $0.2<g-r<0.8$ and
$0.6<u-g<2.2$ are selected. The $u$ magnitude error is much larger
than that of $g$ and $r$, especially at the faint end.}
\label{figure1}
\end{figure}

 The SDSS is a large international collaboration project,
and it has obtained deep, multi-color images covering more than
one-quarter of the celestial sphere in the North Galactic cap, as
well as a small ($\sim 300$ deg$^{2}$), but much deeper survey, in the
South Galactic hemisphere \citep[][]{York00}. The SDSS used a dedicated 2.5-meter
telescope at Apache Point Observatory, New Mexico \citep[][]{Gunn06}. The flux
densities are measured in five bands ($u$, $g$, $r$, $i$, $z$) with
effective wavelengths of 3551, 4686, 6165, 7481 and 8931 \AA,
respectively.  The $95\%$ completeness limits of the images are
22.0, 22.2, 22.2, 21.3, and 20.5 for $u$, $g$, $r$, $i$, and $z$,
respectively \citep[][]{Abazajian04}. The relative photometric
calibration accuracy for $u$, $g$, $r$, $i$, and $z$ are 2\%, 1\%,
1\%, 1\% and 1\%, respectively \citep[][]{Padmanabhan08}.
Figure \ref{figure1} shows the error of $u$-, $g$-, and $r$-band
magnitude as a function of $g$-band magnitudes. Other technical
details about SDSS can be found on the SDSS website
\emph{http://www.sdss3.org/}, which also provides an interface for the
public data access.

\par The Sloan Extension for Galactic Understanding and Exploration (SEGUE-1) obtained spectra of nearly 230,000
unique stars over a range of spectral types to investigate
Galactic structure. Building on this success, SEGUE-2
spectroscopically observed around 119,000 unique stars, focusing on
 the stellar halo of the Galaxy, including stars with distance from 10 to 60 kpc from the Sun.
 We employ the complete set of derived stellar parameters for SEGUE-1 and SEGUE-2 from the
newest version of SEGUE Stellar Parameter Pipeline
\citep[SSPP;][]{Beers06, Lee08a, Lee08b, Allende08, Smolinski11, Lee11}, including effective temperature, surface gravity,
and metallicity (parameterized as [Fe/H]). A thorough overview of SEGUE effort can
be found in \cite{Yanny09}.

\par We use the adopted stellar atmospheric parameters from the SSPP listed in the sppParams table. After
excluding the repeated stars surveyed on different plates, we  obtain a sample of 366,676 stars with SDSS $u,g,r,i, and~z$
magnitudes, as well as stellar parameters. Most stars in
the sample have metallicities in the range $-2.5\le$[Fe/H]$\le 0.0$. In this
study, in order to generate the [Fe/H] probability distribution from the colors,
we select main-sequence stars, adopting the following selection criteria:
\begin{itemize}
\item $14<g<19.5$;
\item $0.2<g-r<0.8$;
\item $0.6<u-g<2.2$;
\item Main-sequence stars are selected by rejecting those objects
at distances larger than $0.15$ mag from the stellar locus described by following equation \citep{Juric08}:
\begin{align}
g-r =& 1.39\{1-exp[-4.9(r-i)^3-2.45(r-i)^2 \nonumber \\
 & -1.68(r-i)-0.05]  \} \nonumber
\end{align}
\item We further refine the selection of  main-sequence stars by rejecting those objects
at distances larger than $0.3$ mag from the stellar locus
described by the following equation \citep{Jia14}:
\begin{align}
u-g=exp[-(g-r)^{2}+2.8(g-r)-1]  \nonumber
\end{align}
\end{itemize}

\begin{figure*}
\includegraphics[width=1.0\hsize]{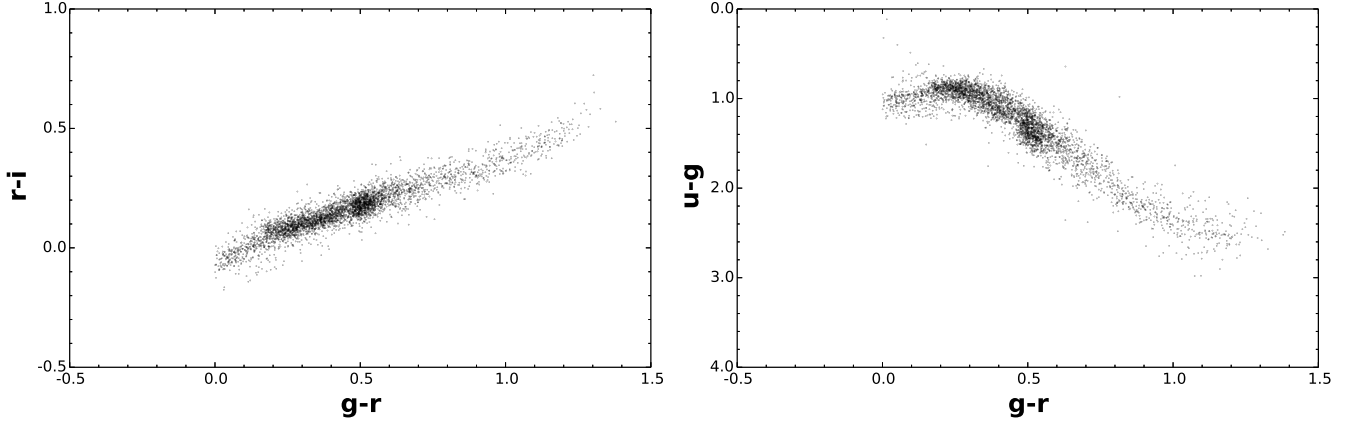}
\caption{Main-sequence stars selection from two-color diagrams. In
the left panel ($r-i$ vs. $g-r$), we reject those objects at
distances larger than $0.15$ mag from the stellar locus. In the
right panel ($u-g$ vs. $g-r$), we further refine the selection of
main-sequence stars by rejecting those objects at distances larger
than $0.3$ mag from the stellar locus.} \label{figure2}
\end{figure*}

As an illustration, Figure \ref{figure2} shows the two-color diagrams for $r-i$ versus
$g-r$ and $u-g$ versus $g-r$. Our final sample includes 268,029 sample stars.
Throughout this paper it is understood that
magnitude and color have been corrected for extinction and reddening following \cite{Schlegel98}.

\begin{figure*}
\includegraphics[width=1.0\hsize]{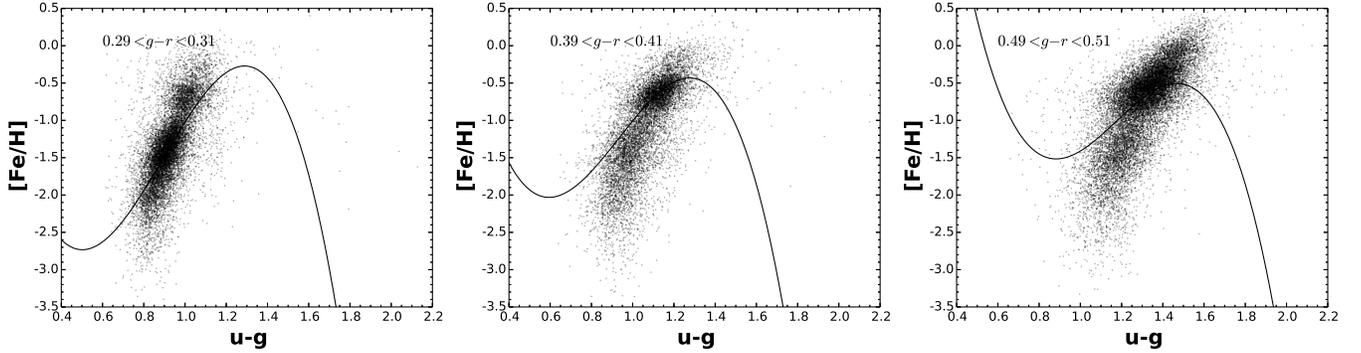}
\caption{The scatter distribution of spectroscopic [Fe/H] versus $u-g$ color for
main-sequence stars with $g-r$ color specified
in a small color interval, as shown in the legends of each panel. The three solid lines
are generated from the photometric metallicity estimator derived by
\cite{Ivezic08}, with $g-r=0.3$ for the left panel, $g-r=0.4$ for the
middle panel, and $g-r=0.5$ for the right panel.}
\label{figure3}
\end{figure*}

\section{Method}

\par Figure \ref{figure3} shows the spectrum-determined [Fe/H] distribution of stars versus
$u-g$ color. These stars are selected within small color ranges
around $g-r=0.3$, $g-r=0.4$ and $g-r=0.5$. The solid lines represent
the photometric metallicity estimator derived by \cite{Ivezic08}
with $g-r$ color specified by 0.3, 0.4, and 0.5 respectively. From inspection of this Figure, these lines can only roughly describe the relation
between [Fe/H] and $u-g$ color. Those stars with specified $u-g$ and
$g-r$ colors exhibit a metallicity distribution which may result
from several other factors. We build the
metallicity probability distribution from the scattered points
in each specified color bin, and then use a monte-carlo method to infer the MDF for large numbers of stars.
That is to say, what we obtain from our photometric calibration is not a one-to-one function, but a probability-governed one-to-many function.

\par The monte-carlo method relies on repeated random sampling to obtain
numerical results. As mentioned in the selection criteria, the two
colors we employ are confined to $0.2<g-r<0.8$ and $0.6<u-g<2.2$. We divided
the $u-g$ vs. $g-r$ panel into $0.05\times 0.05$ mag$^2$ bins, and
designate each square unit by an index computed in the
following manner:
\begin{align}
index=int((u-g-0.6)/0.05)*12+int((g-r-0.2)/0.05)\text{,} \nonumber
\end{align}
where the
symbol $int$ stands for the integer portion. We obtain 384
square units. For a given sample of stars, we also divided the value
of [Fe/H] from -3.5 to 0.5 into 80 bins equally, with a bin width of 0.05
dex. In this manner, we obtained an array of index $\times$ [Fe/H]. Each array
element records the number of stars whose colors and
[Fe/H] match their corresponding positions, based on the above-selected sample of 268,029 stars. The resulting array, with its
array elements holding the pertinent information, is called the ``seed'' array.
Each element of the seed array is denoted by $value[index][i]$, where $i$
ranges from 0 to 79. The maximum of $value[index][i]$ for $i$ in the
range $[0, 79]$ can easily be found, and is denoted by $max[index]$. From this seed array, we can evaluate the MDF for the
photometrically-surveyed stars.

\par For a certain index ($u-g$ and $g-r$ interval),
we use the monte-carlo method to generate a random number sequence
whose probability distribution is determined by the [Fe/H]
distribution recorded in the seed array. Suppose $X$ and $Y$ are two
stochastic variables which can be assigned a value by a random
number generating function $rand_X()$ and $rand_Y()$, respectively.
The function $rand_X()$ is modulated to generate a uniform-probability
distributed random integer number from 0 to 79, and $rand_Y()$ from
0 to $max[index]$. In each trial, we obtain a random number pair
($X=rand_X()$, $Y=rand_Y()$). When $Y\le value[index][X]$, we record
$X$ as a useful value, otherwise we discard it. By numerous trials,
we obtain a sequence of random numbers
\verb"{"$X_{1},X_{2},X_{3},\cdots$\verb"}" that follow the same
probability distribution as those recorded in the seed array. Then we
can transform the obtained random number sequence into
metallicities, [Fe/H], by [Fe/H]=$0.05*index-3.5+0.025$. Here,
because $value[index][i]$ can be equal to zero for some $i$ values,
we can discard them and record the non-zero elements and their
positions in a new array. Through this trick, we greatly improve the
sampling efficiency.

\par For a large number of photometrically-surveyed stars, we first count the number of stars that corresponds to each array element,
and then, following the above procedure, obtain metallicities of the counted number for all array elements. The metallicities naturally
form a distribution which we discuss below.

\begin{figure*}
\includegraphics[width=1.0\hsize]{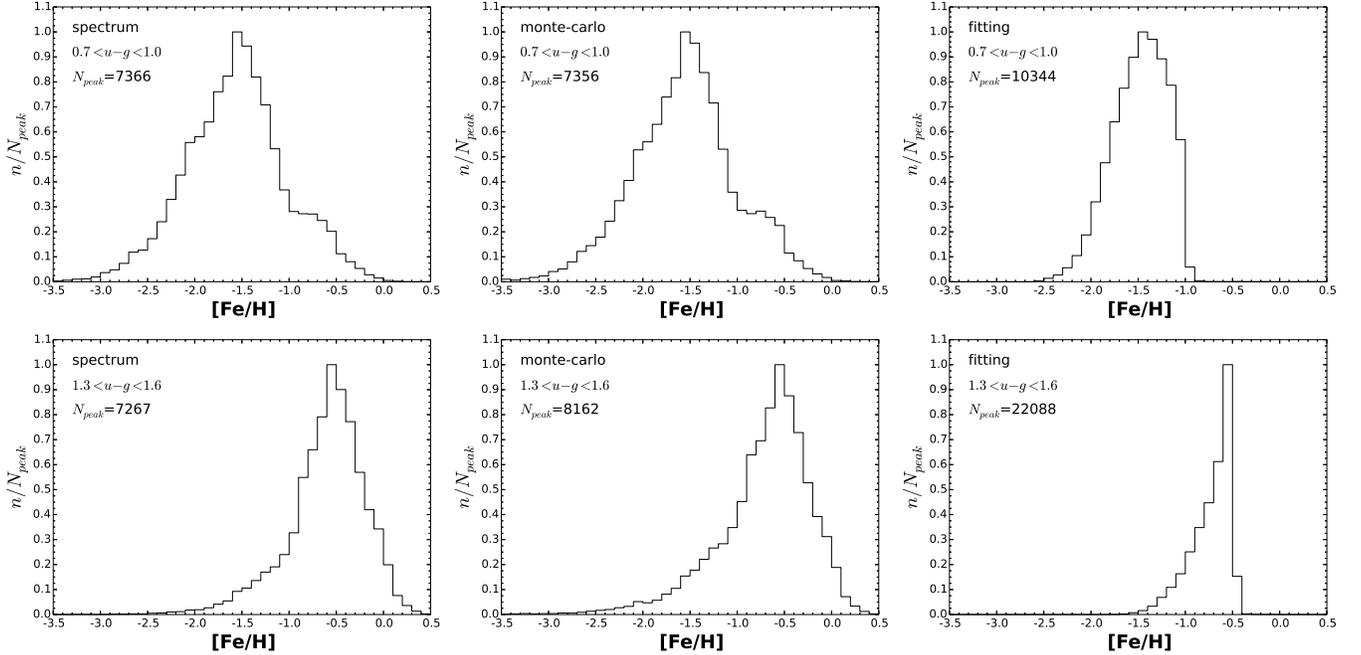}
\caption{The metallicity distribution of spectroscopically-surveyed
main-sequence stars in the color range $0.2<g-r<0.6$. The top three panels show the metallicity
distribution of stars with $0.7<u-g<1.0$, and the bottom
three with $1.3<u-g<1.6$. The left panels present the spectrum-based
metallicity distribution. The middle panels show the photometric
metallicity distribution determined by the monte-carlo method, and the
right panels show the photometric metallicity distribution from
\cite{Ivezic08}'s model. The peak values in all the histograms are
normalized to one, with the actual values shown in the legends of each panel.}
\label{figure4}
\end{figure*}

\section{Comparison}

\par To test the feasibility, and to show the superiority of the method described above,
we make a comparison between this monte-carlo based method and the
third-order polynomial-fitting method presented in \cite{Ivezic08}, shown below:
\begin{align}
\rm[Fe/H]=&-4.37-8.56x+15.5y+23.5x^2 \nonumber \\
&-39.0xy+20.5y^2-10.1x^3 \nonumber \\
&+12.1x^2y+7.33xy^2-21.4y^3 \text{,}\nonumber
\end{align}
where $x=u-g$ for $(g-r)<0.4$ and $x=(u-g)-2(g-r)+0.8$ for
$(g-r)>0.4$, $y=g-r$.

\begin{figure}
\includegraphics[width=1.0\hsize]{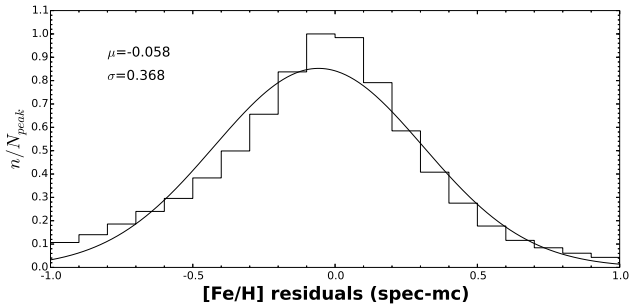}
\caption{The distribution of the [Fe/H] residuals between the
spectrum-based metallicity and that derived by the monte-carlo
method for individual stars. Main-sequence stars are selected with
$0.2<g-r<0.6$. This distribution is fitted by a Gaussian, with its
mean zero-point offset and dispersion values labeled.}
\label{figure5}
\end{figure}

\par We select main-sequence stars covering the color range $0.2<g-r<0.6$
from the spectroscopically-surveyed stars and evaluate their
photometric metallicity distribution. Figure \ref{figure4} shows the
derived MDFs from the three different methods. The top three panels
show the MDFs for stars with $0.7<u-g<1.0$, and the bottom three
panels with $1.3<u-g<1.6$. The left panels present the
spectrum-based MDF which can be regarded as ``ground truth''. The
middle panels show the photometric MDF determined by our monte-carlo
method, and the right panels show the photometric MDF from
\cite{Ivezic08}'s model. The peak values in all the histograms are
normalized to one, with the actual peak values labeled. As shown in
Figure \ref{figure4}, the photometric MDFs based on our monte-carlo
method much more closely resemble the spectroscopic MDFs than those
obtained from the \cite{Ivezic08} approach. Not only are the peak values
of the photometric MDFs determined by our monte-carlo method
approximately equal to those of the spectrum-based MDF, but the
wings of the two distributions have almost the same profile. By
contrast, there exists some clear discrepancies between the
photometric MDFs from \cite{Ivezic08}'s model and the spectrum-based
MDFs, particularly at the very metal-rich and very metal-poor ends.
We thus believe that polynomial-based fitting methods are
clearly inferior, and should no longer be generally used. For
application to a large number of stars, there is great advantage in
using the monte-carlo method for the derivation of photometric MDFs.
Note that if the number of stars used to evaluate the photometric
MDFs is small, we can easily increase the number of desired random
numbers $X$ using a multiplicative factor in the monte-carlo method.

\par Figure \ref{figure5} shows the distribution of
metallicity residuals between the spectrum-based metallicity and that
determined by the monte-carlo method for individual stars. We determine the metallicity of a single star from the
peak value of the metallicity distribution in each specific color bin.
The stars are selected covering the color range $0.2<g-r<0.6$. The distribution
is roughly a Gaussian profile, with mean zero-point offset of $-0.058$ dex and
dispersion of $0.368$ dex.

\par In the conventional photometric metallicity calibration by polynomial fitting, the error mainly arises from two sources. One is
from the fitting method itself (illustrated in Figure \ref{figure3}), and the other is from the errors
in the color indexes. In SDSS, the photometric error of the $u$-band
magnitude is relatively large (shown in Figure \ref{figure1}), limiting the application range of the photometric metallicity
estimator. In the monte-carlo calibration method, we suppose that
the spectrum-based metallicity distribution in a given $u-g$, $g-r$ color bin
is fairly reliable, and thus, by reproducing its distribution we
effectively eliminate the errors arising from the fitting method. The error
introduced by fluctuations in the monte-carlo method can be
eliminated by increasing the number of desired random numbers. In
addition, the third-order polynomial model is determined by only 10
coefficients, while the model (seed array) construction involves many more variables. This, to some extent,
guarantees the accuracy of photometric metallicity distribution
determined by the monte-carlo method. The deviation of the photometric
MDF determined by the monte-carlo method mainly
arises from the errors in the color indexes, especially when estimating
the MDFs for faint stars. As shown in
Figure \ref{figure4}, the calibration based on the monte-carlo method can be used to derive
photometric MDFs with sufficient
accuracy for most purposes.

\begin{figure*}
\includegraphics[width=1.0\hsize]{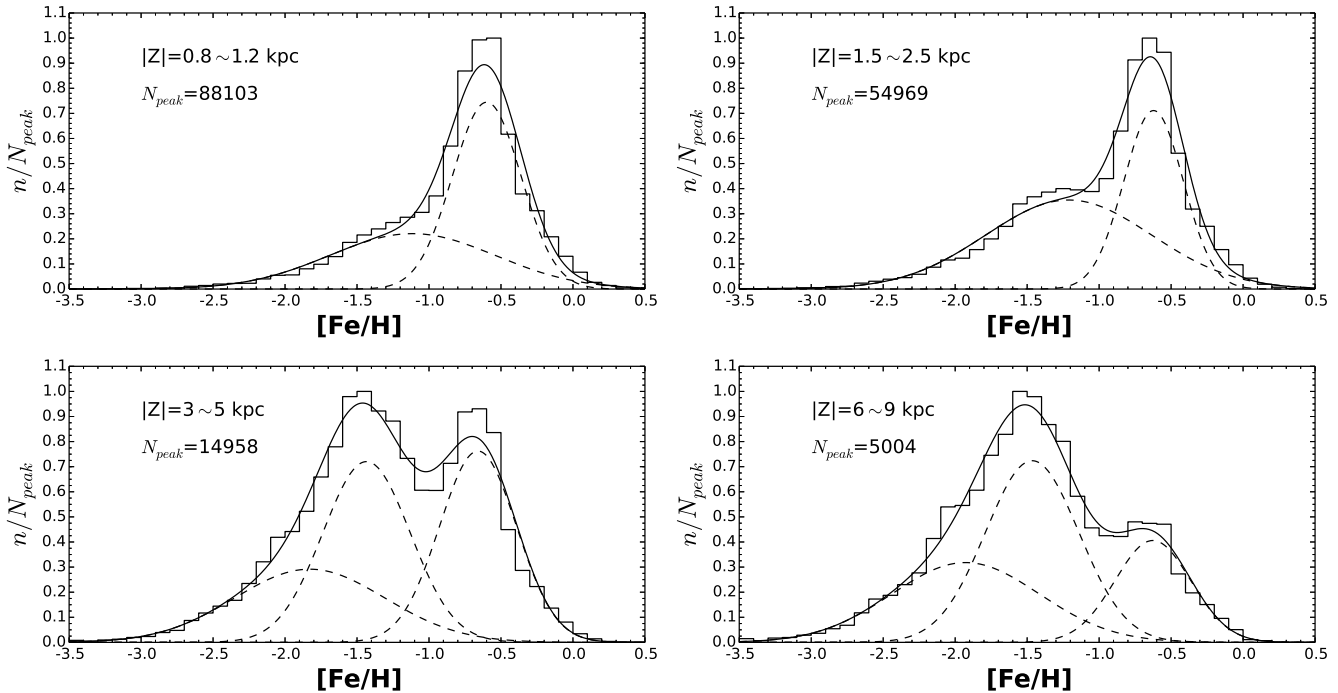}
\caption{Photometric MDFs derived by the
monte-carlo method. Main-sequence stars with $0.2<g-r<0.6$, 6
kpc$<R<$9 kpc, and distance from the Galactic plane in the range
$0.8\sim 1.2$ kpc (top left panel), $1.5\sim 2.5$ kpc (top right
panel), $3\sim 5$ kpc (bottom left panel), $6\sim 9$ kpc (bottom
right panel) are selected as samples. The peak values in these four
histograms are all normalized to one, with actural values labeled. The top two MDFs can be well-fit
by two Gaussians with peaks at $-0.6$ (disk contribution) and
$-1.2$ (halo contribution) respectively. The bottom two MDFs are better-fit by three Gaussians,
with peaks respectively at $-0.6$, $-1.4$ and $-1.9$.}
\label{figure6}
\end{figure*}

\section{Application}
\par As an example, we use the method introduced in this study to derive
the photometric MDF for stars as a function of distance from the
Galactic plane. The main-sequence stars with $0.2<g-r<0.6$, $6$ kpc
$<R<9$ kpc, and in different $|Z|$ intervals are selected ($R$, $Z$
represent cylindrical Galactocentric coordinates, with the Sun's
coordinate $(R,~ Z,~ \phi)=$(8 kpc, 0, 0)). As shown in
Figure \ref{figure6}, the photometric MDFs in the top two panels
can be well-fit by two Gaussians, with peaks at about $-0.6$ and
$-1.2$ respectively, one associated with the disk system, and
the other with the halo (and metal-weak disk). The MDFs in the two bottom
panels are found to be better-fit by three
Gaussians, with peaks at about $-0.6$, $-1.4$ and $-1.9$, respectively.
The two lower peaks may be associated with
inner-halo and outer-halo populations, respectively \citep[][]{Carollo07, Carollo10,An13,
An15}. These four histograms clearly show that the number ratio
between disk stars and halo stars decreases with vertical distance
from the Galactic plane. In the metal-rich and metal-poor ends, the
number of stars decrease gradually. The number ratios as a function of
$|Z|$ between the disk and halo above the Galactic plane could be
recalculated from the photometric metallicity distribution, and the
result can be compared with that from the method of star counting.
This will be presented in our future papers.

\section{Summary}

\par This paper presents a new method to estimate the photometric MDF
for main-sequence stars. The method is tested using a spectroscopic
sample of stars from the SDSS. Compared with the method from
\cite{Ivezic08}, the current method is more accurate, particularly
for very metal-rich and very metal-poor stars. This method is
sufficiently accurate to be used to investigate the distribution and
chemical structure of the Galactic stellar populations.  At the same
time, as an example, we also apply the method to the main-sequence
stars with $0.2<g-r<0.6$, $6$ kpc $<R<9$ kpc, and different $|Z|$
intervals. The metallicity distribution of sample stars near
the Galactic plane can be well-fit by two Gaussians, with peaks at
about $-0.6$ and $-1.2$, respectively, one associated with the disk system and the other with the halo/metal-weak thick disk. However,
the metallicity distribution of the sample stars far from the
Galactic plane can be well-fit by three Gaussians, with peaks
at $-0.6$, $-1.4$ and $-1.9$, which supports the
existence of two components in the halo: the inner-halo and the outer-halo.
The number ratio between disk stars and halo stars varies with
vertical distance from the Galactic plane. In the metal-rich and
metal-poor ends, the number of stars decreases gradually. With the
advantage of the method introduced in this paper, we can better
study the photometric MDF for different stellar populations in the
Galaxy and provide detailed constraints on the Galactic chemical
evolution, which we will consider in future papers.

\section*{Acknowledgements}

\par We especially thank the referee for his/her insightful comments and suggestions
which have improved the paper significantly.
This work was supported by joint fund of Astronomy of
the National Natural Science Foundation of China and the Chinese
Academy of Science, under Grants U1231113.  This work was also by
supported by the Special funds of cooperation between the Institute
and the University of the Chinese Academy of Sciences. In addition,
this work was supported by the National Natural Foundation of China
(NSFC, No.11373033, No.11373035), and by the National Basic Research
Program of China (973 Program) (No. 2014CB845702, No.2014CB845704,
No.2013CB834902).

Funding for SDSS-III has been provided by the Alfred P. Sloan
Foundation, the Participating Institutions, the National Science
Foundation, and the U.S. Department of Energy Office of Science. The
SDSS-III web site is \emph{http://www.sdss3.org/}. SDSS-III is
managed by the Astrophysical Research Consortium for the
Participating Institutions of the SDSS-III Collaboration including
the University of Arizona, the Brazilian Participation Group,
Brookhaven National Laboratory, Carnegie Mellon University,
University of Florida, the French Participation Group, the German
Participation Group, Harvard University, the Institute de
Astrofisica de Canarias, the Michigan State/Notre Dame/JINA
Participation Group, Johns Hopkins University, Lawrence Berkeley
National Laboratory, Max Planck Institute for Astrophysics, Max
Planck Institute for Extraterrestrial Physics, New Mexico State
University, New York University, Ohio State University, Pennsylvania
State University, University of Portsmouth, Princeton University,
the Spanish Participation Group, University of Tokyo, University of
Utah, Vanderbilt University, University of Virginia, University of
Washington, and Yale University.

\end{document}